\documentclass[11pt,letterpaper]{article}
\usepackage{amsmath}
\usepackage{amssymb}
\usepackage{graphicx} %package from Garching paper
\usepackage{natbib}
\usepackage{enumitem}
\usepackage{caption}  %[caption=false]{caption}  %'d out when chgd to aastex 2013-05-06
\usepackage{subcaption}  %[caption=false]{subcaption}  %'d out when chgd to aastex 2013-05-06
\usepackage[caption=false]{subfig}  %\usepackage{subfig}  %\usepackage[subfigure]{tocloft}
\usepackage{authblk} %[affil-it]{authblk}%'d out when chgd to aastex 2013-05-06
\usepackage{mathabx}  %'d out when chgd to aastex 2013-05-06
\usepackage{deluxetable}%{•}

\setlist{nolistsep} %,leftmargin=*}

\ifx\UseOption\undefined
\def\UseOption{optb}
\fi
\usepackage{optional}

\begin{document}
%\newcommand{\qprimeast0}{$Q^{\prime}_{\ast}$}
%\newcommand{\qprimeast1}{$1/Q^{\prime}_{\ast}$}

%\title{Signs of planet loss indicate ongoing orbital instability of main sequence exoplanetary systems}
%\setcounter{page}{1}
%\include{NASAdata}%.tex}

%\include{NASAsumm} %.tex}
%\setcounter{page}{1}
 %\setcounter{section}{1}
 %\include{NASAcontents}
 %\setcounter{page}{1}
  %\setcounter{section}{0}
%\include{FeEccFlowApJ_ProvSummary}   

% ~/Dropbox/talks_db/2013planetloss/  ...  HotBigPlanetsKepler_SolicitationCover.tex
% c:/Users/Stuart/Dropbox/talks_db/2013planetloss/
% From % ~/Dropbox/talks_db/2013planetloss/2013feEccCorrPaper/FeEccFlowApJ_Text.tex

\begin{center}
\begin{LARGE}

Seeking Support For My ``Campaign For Participation'' 

\end{LARGE}
\end{center}

%\section{Summary of Presentation On: Iron Correlations, Distribution Discrepancy, and Planet Loss }
%\label{sec:presummary}                

I submit this paper  as part of my
``Campaign for Participation''
in exoplanet science. It is submitted in response to NASA's call for white papers on alternative 
science that can be done by Kepler. My purpose is to make a worthy contribution to help obtain collaborators
and to get attention for my job search, 
all to remain a part of the collaborative study of exoplanets.

This work makes use of the statistics from Kepler and ground searches to show that the greater numbers of giant planets at the shortest periods represents an excess given that there are more medium planets in general, and given that giant planet should migrate into the star faster. I present here and in Taylor 2013b that this is evidence of ongoing giant planet migration.
It is essential that the community help credit me for being a part of the work obtaining these statistics in order that I have the paper credit to remain in the field by returning to full employment. I am campaigning to take back my chance to continue the LCOGT transit confirmation that I started by putting my missing contributions back into the papers. Please collaborate with me when you collaborate with this UCSB-affiliated observatory to make sure that I get a chance to be part of work that I started. I have done everything possible to go around the observatory trying to keep me out, but it necessary for  outside collaborators to include me because staff scientists are not allowed to talk to me. I am not allowed to apply for observing.  Despite having been the onsite astronomer for the FTN telescope, I have been allowed on no papers giving results from my work. 
My job search shows it has been essential for me to have this credit to obtain new employment. 
I hope  authors using LCOGT planet-confirming observations will seek to have an honest list of authors. 
My unending persistence to interpret these statistics shows the most important reason to include me: I \textit{want to be a part} of learning about exoplanets.

Please help me get back onto where I was going as an exoplanet astronomer. I appeal to those in the community to whom I apply for employment, to please consider me to have contributed to several UCSB/LCOGT papers at an author level. I am doing everything I can to be a part of the projects I started, but have been given no chance to finish what I started. The upcoming publication of the LCOGT global network paper (submitted to PASP) is an example of a paper that I contributed to at an author level, but that the observatory is keeping me out of despite years of effort to be included. I call for standards requiring giving further participation to all contributors. I place here another record that I have done all I can to be on many papers but am allowed no chance.

I also appeal to  philanthropists for support to continue this work as an astronomer. I seek support to meet basic needs and to be able to start a family.

I am grateful for messages of support. I solicit recommendations that I would be allowed to use in my job search. Besides replacing old letters of recommendation, I could also cite approval to make up for how white papers are not given peer review. This is my work alone, but I hope to be working with others soon.

\setcounter{page}{1}
\setcounter{section}{0}
% ~/Dropbox/talks_db/2013planetloss/2013KeplerWhitePaper/HotBigPlanetsKepler_Text.tex        
%HotBigPlanetsKepler_Text.tex
% From ~/Dropbox/talks_db/2013planetloss/2013feEccCorrPaper/FeEccFlowApJ_Text.tex
% c:/Users/Stuart/Dropbox/talks_db/2013planetloss/2013feEccCorrPaper/FeEccFlowApJ_Text.tex
% From ~/Dropbox/jsrch_db/cv+ri_jsrch_dbox/0researchInterestsLatex/ 

%\title{Signs of planet loss indicate ongoing orbital instability of main sequence exoplanetary systems}
\title{Hot Big Planets Kepler Survey:  Measuring the Repopulation Rate of the Shortest-Period Planets} 
%\title{Hot Big Planets Mission: Studying Tidal Migration Through Measuring Occurrence Distribution % indicate 
%of Large and Medium Planets}                      % in %orbital instability of 
%\title{Iron abundance correlations and an occurrence distribution discrepancy as results of % indicate 
%ongoing inward planet migration in %orbital instability of 
%main sequence exoplanetary systems}
% 
\author{Stuart F. Taylor}
\affil{Participation Worldscope \\ Sedona, Arizona, USA \\ astrostuart@gmail.com}
\date{\vspace{-5ex}} %{-10ex}}
\maketitle

%\begin{itemize}
%\opt{opta}{\item First item}
%\opt{optb}{\item Second item}
%\end{itemize}

\begin{abstract}

By surveying  new fields for the shortest-period ``big'' planets, the Kepler spacecraft
could provide the statistics to more clearly measure the occurrence distributions of giant and medium planets.
This would allow separate determinations for giant and medium planets 
of the relationship between 
the inward rate of tidal migration of planets
and the strength of the stellar tidal dissipation (as expressed by the tidal quality factor $Q^{\prime}_{\ast}$. 
We propose a ``Hot Big Planets Survey'' to find new big planets
to better determine the planet occurrence distribution
at the shortest period. 
We call planets that Kepler will be able to find ``big'',
for the purpose of comparing the distribution of giant and medium planets
(above and below 8 earth radii).

The distribution of planets from one field has been interpreted to 
show that the shortest period giant planets are at the end of an ongoing flow of high eccentricity migration,
likely from scattering from further out. 
The numbers of planets at these short periods is still small, leaving uncertainty 
over the result that the distribution shows the expected power index for inward tidal migration.
The current statistics make it hard to say whether the presence of more giant planets at the shortest periods 
despite there being more medium planets at most periods 
indicates a greater migration of giant than medium planets.

We propose a repurposed Kepler mission to make  45-day observations to survey
10 times as many stars as in the survey of the original field, to
survey for planets with periods of up to fifteen days with at least three transits.
This would better measure the occurrence distribution in the period range that  includes planets migrating into 
the star, allowing a better measure of the relative rate of migration of giant and medium planets.
Observations of several Kepler fields would give  the statistics to better determine 
important parameters of planet migration and tidal strengths.

\end{abstract}
%Keywords: planetary systems – techniques: photometric – techniques, 
%planetary systems: dynamical evolution, tidal interactions,
%binaries (including multiple): close, planetary systems: formation, stars: fundamental parameters
%Planet-star interactions

\section{Introduction}
\label{sec:intro}        
The Kepler mission has provided a wealth of statistics %and ground based planet discoveries have
on the occurrence distribution of planets. Statistics from the Kepler mission are much more
reliable than statistics from ground based surveys due to difficulty in understanding the
day/night window function of transit surveys and the varying systematics of all ground surveys.     
Now that the Kepler spacecraft will have less photometric precision due to operating with 
only two reaction wheels, NASA has called for submissions of alternative mission concepts.
The spacecraft will be unable to continue pursuing its original goal of observing 
transits of planets of radii the order of the earth in year-long orbits.
We propose that Kepler do what transit surveys  do best: Find the highest number
of ``big'' radii planets as possible by observing many fields. 
We call this the ``Hot Big Planets Mission'', where we define ``big'' to include from the largest planets
down to whatever radii Kepler is still able to detect. 
We refer to planet candidates as ``planets'', even for objects 
possibly too large to be planets. %(even \textgreater 16 $R_{\oplus}$).
 % as ``planets'' even though their planetary nature is uncertain.
Though we hope Kepler will still be
able to detect smaller planets, our purpose here is to show the value of comparing 
the occurrence distribution of planets above and below 8 earth radii ($R_{\oplus}$).

The measurement of the shortest period occurrence distribution obtained using 
data from one Kepler field has shown  tantalizing patterns that are still unresolved 
by the available statistics. 
Results from ground surveys and from 
surveying the first Kepler field show a dropoff in the distribution that can
be attributed to stellar tides causing planets to undergo inward migration into the star, 
but currently available statistics are still low, 
producing only very uncertain measurements of the parameters of the innermost distribution.
However, we are  left with the mystery of how there are
some planets so close to the star that it is improbable to have found them
in the last short time period before their merger with the star, as would be expected if the tidal migration 
is similar to that measured for binary stars 
\citep{ham09,heb10,bir13,mei05}.
The shortest period occurrence distribution of giant planets has lead to work concluding 
that tidal dissipation on stars caused by tides raised by planet-mass companions
is unexpectedly weak \citep{pen12}. 
The number of giant planets at the shortest periods is in relative excess to the number of
medium planets, and the distribution of planets from 4 to 8 $R_{\oplus}$ can be explained
with a stellar tidal migration strength not excessively weaker than that measured for binary 
stars. This has been used by 
\citet[hereafter T12b,T13b]{tay12b,tay13b} 
to show that the occurrence distribution of could be explained by the
migration of giant planets (``flow'') proposed by \citet{soc12} which could repopulate 
the shortest period occurrence distribution
as planets are lost to inward migration due to tides on the star. With such a flow,
the tides on stars need not be weaker for 
planetary mass companions than they are for stellar mass companions. %\citep{soc12,tay12a,tay13b}. 
Though Kepler has found many planet candidates, 
the observation of one field has not provided enough of the closest planets to 
show the planet occurrence distribution as a function of radii clearly enough to
separate out the effects of tidal migration from changes (up or down) in planet radii.
It is easy to forget that the shortest period planets are actually quite rare -- a fact obscured by how 
transit surveys find the closest planets much  more easily.
Kepler will find more planets faster if it observes for 45 days and then
moves to a new field. By observing many fields, it will provide important statistics that 
will show what happens as planets tidally migrate into their stars.

The one-field Kepler data has been interpreted to show that giant and medium planets 
undergo tidal migration when they get too close to the star 
(T12b,T13b, \citet[Taylor 2012a, Taylor 2013a; hereafter T12a,T13a]{tay12a,tay13a}), %{tay12a,tay12b,tay13a,tay13b}, 
but the uncertainty in the fits 
\citep[hereafter H12]{how12} 
are too high to be certain we see current inward migration rather than 
the distribution left over when the protoplanetary disk dissipated. 
However, the small numbers of planets leaves noise that obscures
whether the patterns have more detail that could help show 
whether other factors could affect the distribution. 
The small numbers of planets from one Kepler field give the distribution insufficient clarity 
to consider how changes in radii could affect the distribution. 
There is reason to believe that planets can inflate or can suffer atmospheric loss. 
True,  mass measurements will also be important to study these problems, 
but there will still be too few planets at short periods to separate out the effects 
of tidal migration and the changes in radii, either up or down, 
that planets can experience as they migrate to merger with the star.

We propose a repurposed Kepler mission to conduct observations of several %repurposed-objective
new fields for periods of 45 days with the objective of better measuring the occurrence
distribution of ``Hot Big Planets'' (HBP), that is, short-period planets of giant and intermediate 
radii. Even if Kepler is only able to monitor half as many stars in each new field as in the orginal field, 
the number of stars surveyed for planets of periods of fifteen days or less would be increased by
a factor of four each year. We hope that five to ten fields could be observed. 
It would likely be possible to still find planets in 15 day periods with three 15 day
observations spaced to find longer period planets. 

We show that there are important questions of the nature of planets, stars, and their
interactions that are best measured by improving the statistics of big planets.
We highlight the differences in the distribution of ``giant'' (\textgreater 8 earth radii, $R_{\oplus}$)
and ``medium'' (4 to 8 $R_{\oplus}$) radii planets in current Kepler data,
but point out that because these shortest period planets are in fact relatively rare
(despite their over-representation in early discoveries) that there remains large
uncertainties in the occurrence distribution of planets at periods of a few days.
There appears to be an excess of giant relative to intermediate planets at the shortest periods,
given that at slightly longer periods, there are more intermediate planets.

\begin{figure}[htb]  %Not needed: [htb]
\centering
  {\includegraphics[width=0.5\textwidth]{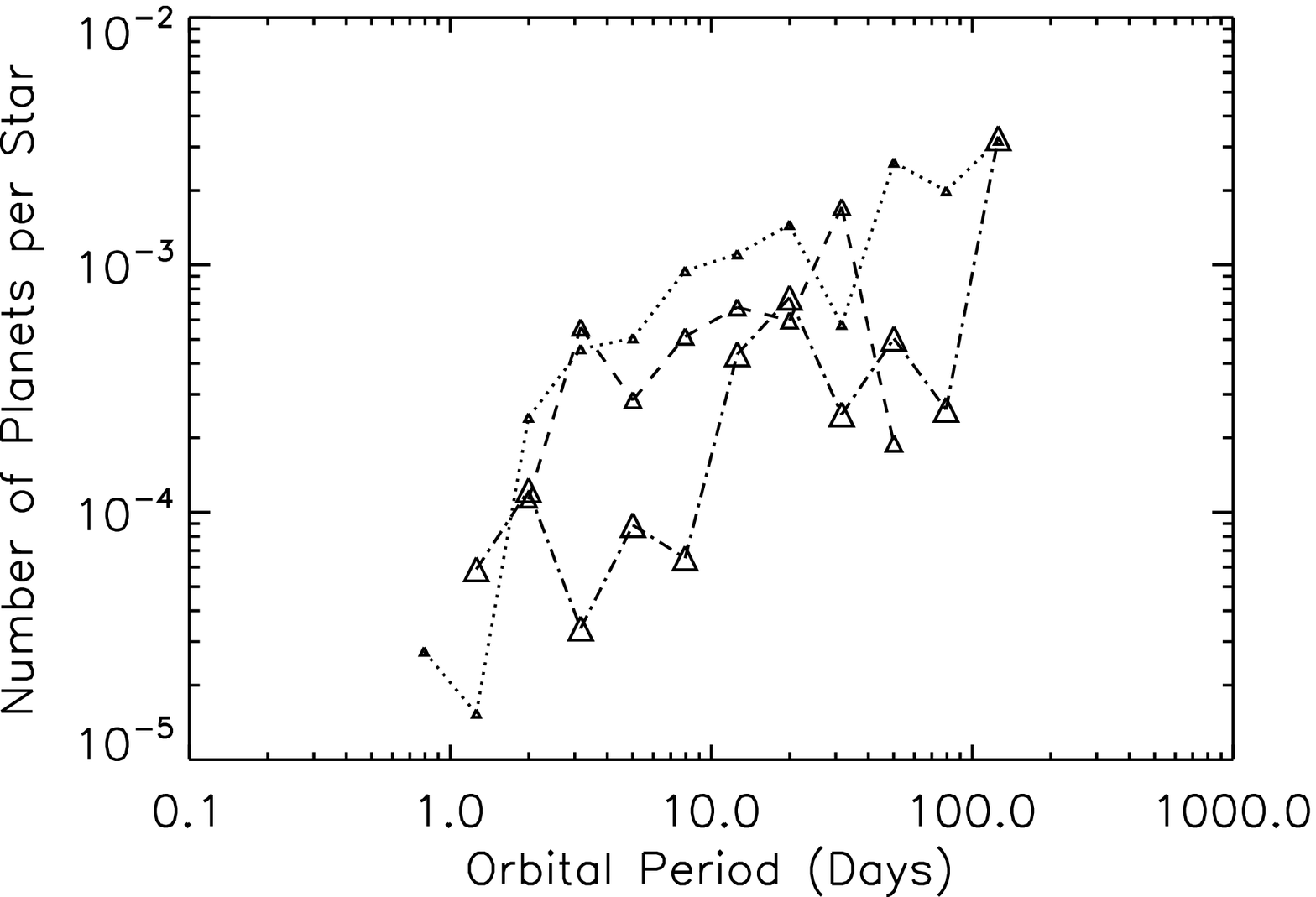}}%.96
\caption{Normalized counts of Kepler ``planet candidates'' above 8 $R_{\oplus}$,
  divided into three radii ranges, where the smallest triangles indicate the smallest
  radii planets, 8 to 11.3 $R_{\oplus}$, the medium range is 11.3 to 16 $R_{\oplus}$,
  and the largest is above 16 $R_{\oplus}$.
 }
  \label{fig:jupdiv3rad}
\end{figure}

It appears hopeful that the two-wheeled mission may still find planets 
sufficiently below 8 $R_{\oplus}$ to characterize the distribution of medium radii planets. 
%We seek to count the fraction of stars with planets above and below 8 $R_{\oplus}$. 
There appears to be a difference in the occurrence rates of planets from 6 to 8 
$R_{\oplus}$ compared to planets larger than 8 $R_{\oplus}$, so we hope that new observations
will be able to reach a precision of 0.36\% 
(and being able to count planets of radii of 5 or 4 $R_{\oplus}$ would be
even better). However, we see in Figure~\ref{fig:jupdiv3rad} that there are  interesting differences 
even in planets with radii from 
8 to 11.3 $R_{\oplus}$, 11.3 to 16 $R_{\oplus}$, and above 16 $R_{\oplus}$.

%We propose that the Kepler spacecraft be used to observe a few new fields to 
%improve the statistics of planets big enough for it to still measure in two-wheel mode.

%We propose a goal of surveying
%ten times as many stars as in the original Kepler field for periods of 30 to 45 days. 

Improved knowledge of the occurrence distribution of ``big planets'' at the shortest period 
would improve understanding of the evolution and properties of interacting planets and stars.
Kepler will still be able to search for ``big'' planets, which we define to be of Jupiter and Neptune
size, which we call ``giant'' and ``medium'' sized planets.
The Kepler mission has already greatly improved measurements of this closest region with its
continuous and unbiased coverage of one field of stars, but the quality of statistics currently available from
Kepler observations are still limited by low number statistics, because planets at the shortest periods
are actually  rare. 
Though in early discoveries of planets, planets with the shortest periods were
overrepresentated due to their relative ease of discovery. However, ground based surveys 
have poorly studied systematics and window functions which makes it difficult to reliably assess their rate of 
discovery, leading to uncertainty in how good the occurrence functions derived from ground surveys
really are.

We propose a goal of observing ten times as many stars as observed in the Kepler field for a period of 45 days.
Even if current limitations and new fields only allow monitoring 75,000 stars at a time, 
half as many stars as the original 150,000, then observing each year of observing new 
fields could still achieve four times the statistics of the original Kepler mission.
Even 30 day observations would improve statistics of 10 day and below planets,
and depending on the reliability of only having two transits, may still be useful for
statistics of planets with up to 15 day periods.
Other than observing new fields for 45 days, the details of this survey will be similar
to the nominal mission. Selection of the fields is beyond one paper, but the first field
could be from the alternate fields considered for the nominal mission.
Finally, it may be possible to break up the observations in 15 day pieces to allow surveying of
longer period planets as well.
This project is highly flexible in choice of fields to allow for other types of
observations.

The new fields could be selected to include a wider range of star ages, with an effort being made to
include younger stars and more stars in clusters, in order to look for changes of the planet occurrence function with age.
This would allow determining whether the difference in the three-day pileup of giant plants 
between the Kepler field and the solar neighborhood is due to a different distribution of stellar ages.

%\subsection{Most interesting features of planet distribution}
%\label{ssec:mostinteresting}                
%Most interesting features of the Kepler planet candidate occurrence distribution:
%\begin{itemize}
%  \item Power index (slope in log-log plot).
%  \item Relative excess of giant versus medium planets.
%  \item Inflated planets – only known when mass is known.
%  \item Short period upturn that could be partially evaporated plants. 
%    Even the theoretical prospect of partially evaporated planets is worth researching.
%  \item Two populations, usually iron rich and iron poor, crowded and uncrowded.
%\end{itemize}

%%%% \section{Evidence for ongoing migration}
%%%% \label{sec:ongoing}
%\subsection{Organization}				%\label{subsec:organization}    
We show how obtaining a clear measurement of the distribution of such ``big'' planets will
allow for determination of the relationship between the rate of migration of planets
and the strength of the tidal dissipation in stars, $1/Q^{\prime}_{\ast}$, in Section~\ref{sec:objective}. 
We discuss the difference in giant versus medium planet distributions as evidence for ongoing migration 
of planets  in Section~\ref{sec:MoreGiants}. %{sec:HotBigPlanets}.

%\section{Hot Big Planets} \label{sec:HotBigPlanets}

\begin{figure}[htb]  %Not needed: [htb]
\centering
  {\includegraphics[width=0.45\textwidth]{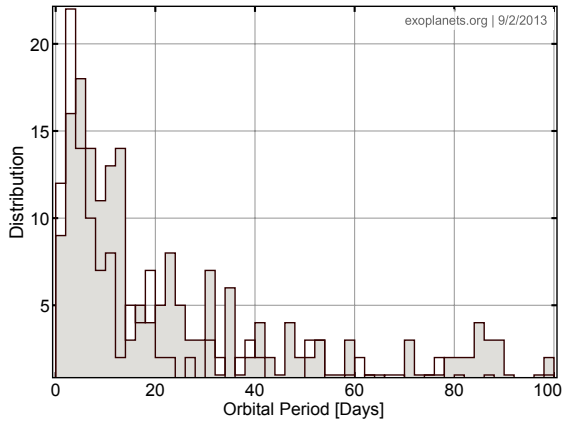}} %.96
\caption{ Raw counts of linear bins of Kepler planet (candidates) with radii
  greater than 4 $R_{\oplus}$, divided into two ranges: 
  The  count of ``giant'' (8 to 16 earth radii, $R_{\oplus}$) are shown by unfilled  bins,
 that of ``medium'' (4 to 8 $R_{\oplus}$) planets are shown by filled bins.  }
  \label{fig:odistrib4rangraw}
\end{figure}

\begin{figure}[htb]  %Not needed: [htb]
\centering
%  \subcaptionbox{All systems.\label{fig:EccVsPerSzFe-All}}
%  {\includegraphics[width=0.48\textwidth]{mass.eps}} %EccVsPerSzFe.eps}}
%  \subcaptionbox{Systemswith \label{fig:EccVsPerSzFe-ltfe0}}
  {\includegraphics[width=0.66\textwidth]{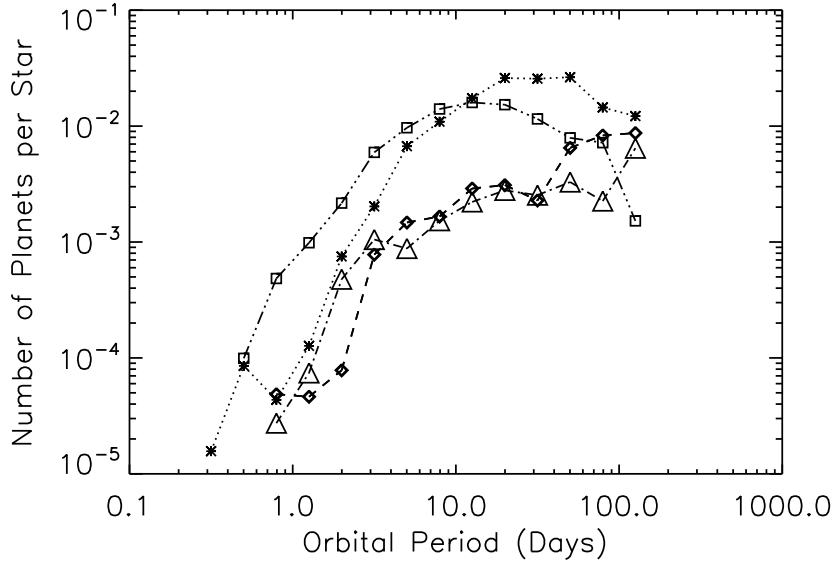}} %.96
\caption{Counts of Kepler ``planet candidates'' normalized by transit probability divided
 by the number of stars, and counted in logarithmic bins
 to give the ``occurrence distribution'' of planets. 
 These normalized counts have been divided into four ranges in radii.
 The normalized count of ``giant'' (8 to 16 earth radii, $R_{\oplus}$) are shown by triangles, %\textgreater 
 that of ``medium'' (4 to 8 $R_{\oplus}$) planets are shown by diamonds,
 that of ``medium-small'' (2 to 4 $R_{\oplus}$) planets are shown by asterisks, and
 that of ``small'' (2 to 4 $R_{\oplus}$) planets are shown by squares.
 }
  \label{fig:odistrib4rangnormed}
\end{figure}

\begin{figure}[htb]  %Not needed: [htb]
\centering
  {\includegraphics[width=0.66\textwidth]{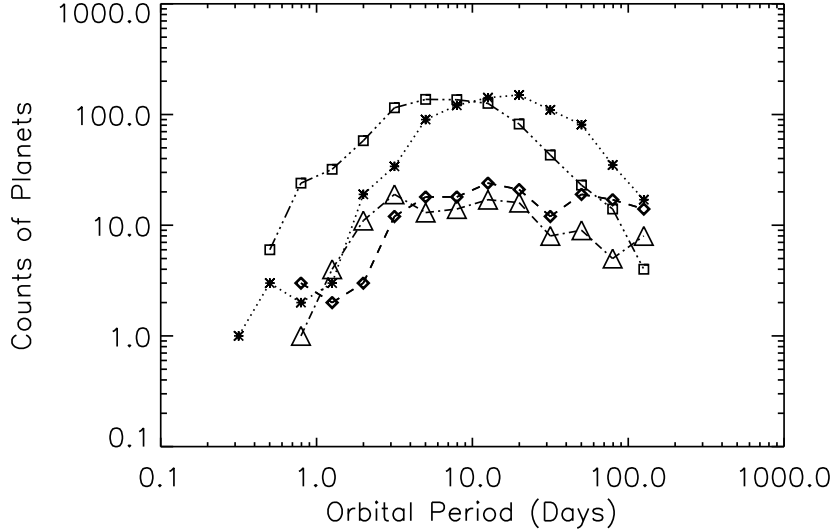}} %.96
\caption{ Raw Counts (not normalized) in logarithmic bins of the observed transiting
 Kepler planets divided into four ranges in radii.
  The normalized count ``giant'' (8 to 16 earth radii, $R_{\oplus}$) are shown by triangles, %\textgreater 
 that of ``medium'' (4 to 8 $R_{\oplus}$) planets are shown by diamonds,
 that of ``medium-small'' (2 to 4 $R_{\oplus}$) planets are shown by asterisks, and
 that of ``small'' (2 to 4 $R_{\oplus}$) planets are shown by squares.
  }
  \label{fig:odistrib4rangraw}
\end{figure}

\section{Objective: Improve Statistics of Big Planets to Study Tidal Migration, Tidal Dissipation, and More}
\label{sec:objective}

Studying planet migration is an important rationale for observing new fields to find new “big” planets. 
%\subsection{Murky patterns}
%\label{subsec:patterns}    
The occurrence distribution for Kepler planet candidates with periods up to 100 days
is shown in Figure~\ref{fig:odistrib4rangnormed}, where the counts of planets has been
normalized by the transit probability and divided by the number of stars observed.
We also show the raw counts in Figure~\ref{fig:odistrib4rangraw} in order to display
how many counts the results in Figure~\ref{fig:odistrib4rangnormed} rely on.
We see important features  in Figure~\ref{fig:odistrib4rangraw},
but comparison with Figure~\ref{fig:odistrib4rangnormed} shows interesting features that
rely on only a few points.
%We would be in a far better position to describe the
%occurrence distribution if we had the points that a HBP mission could provide.

We highlight three features of the shortest period planet distribution:
\begin{itemize}
  \item The falloff of the giant and medium planets has a power index close to
  that predicted for tidal migration into the star.
  \item Though the number of giant planets is generally less than the number of
  medium planets for periods greater than a few days, there is a relative excess of
  giant planets going from the pileup at three day periods down to periods on the
  order of a day.
  \item The falloff of medium and medium small planets appears to flatten out at 
  periods less than two days for medium planets and less than one day for medium
  small planets.
\end{itemize}
Each of these features, when compared to Figure~\ref{fig:odistrib4rangnormed}, 
depends on a number of planets too small to definitively make these conclusions.
It is in these few day range that details of tidal migration as well as details of
the inflation of planets or the blowoff of planet atmospheres can be expected to
become visible.

%The shortest period distribution of planets above a planet-mass threshold is not determined by %giant and medium 
%%the original 
%This threshold is dependent on the strength of the stellar tidal dissipation, but we
%will roughly state that the current threshold for giant and medium planets given 
%Several authors have favored the weak stellar tidal dissipation explanation for the presence of the closest planets.
%\citet{ham09} and \citet{hel09} 
%point out that invoking chance scattering to explain the presence of a 
%planet was unlikely. 
%
%The problem with the first explanation is
%that the position of the intermediate mass planets is at longer periods (a further distance from the star).
%This distribution can be explained with a reasonably weak tidal dissipation.
%Furthermore, it is expected that giant planets will undergo planet-planet scatter more frequently than
%intermediate planets. 
%
%Though the explanation of a ``flow'' of inwardly migrating giant planets can fit the data,
%the uncertainty in the H12 fits is much too large to be sure.
%We propose improving these statistics.

%The occurrence distribution for
%superearths appears to be below this threshold. It has a power index
%not consistent with tidal infal, meaning it may still be  shaped by the primoridial distribution, though
%it may include planets at the end of their high eccentricity migration.

The shortest period distribution has three key parameters: the power index of the fall off, 
the location of the fall off, and the power index of the distribution beyond the fall off (H12).
%Besides the power index, the other key parameter of the shortest period power distribution
%is how short of period the distribution has (how close to the star are the closest planets).

%Several authors have noticed that the distribution has ``too many'' closest planets.
%That is, the location of the fall off of giant planets is at too short of a period.
%Given the expected strength of stellar tidal dissipation, it is statistically improbable to find
%planets so shortly before being destroyed by merger with their star.
%%the shortest period planets. It is improbably that we would find planets with such little time left
%%before they were migrated into merger with the star by stellar tides. 
%There are two ways out of this conundrum:
%1. Stellar tides are much weaker than thought.
%2. New planets are continually migrating inwards.

An HBP survey would provide enough statistics to know the power index and position of the falloff
well enough to determine whether the falloff of giant planets being at a shorter period
than for medium planets is due to giant planet migration, or due to the nature of tidal dissipation in the star.

It is also desirable to look for whether the distribution changes with the age of the system, 
or with changes in other parameters such as stellar mass,
but there needs to be more planets to be able to subdivide these statistics by age.

%\subsection{Stellar tidal dissipation}
%\label{subsec:stellarQ}    

\begin{figure}
\centering
\includegraphics[width=0.8\textwidth,clip]{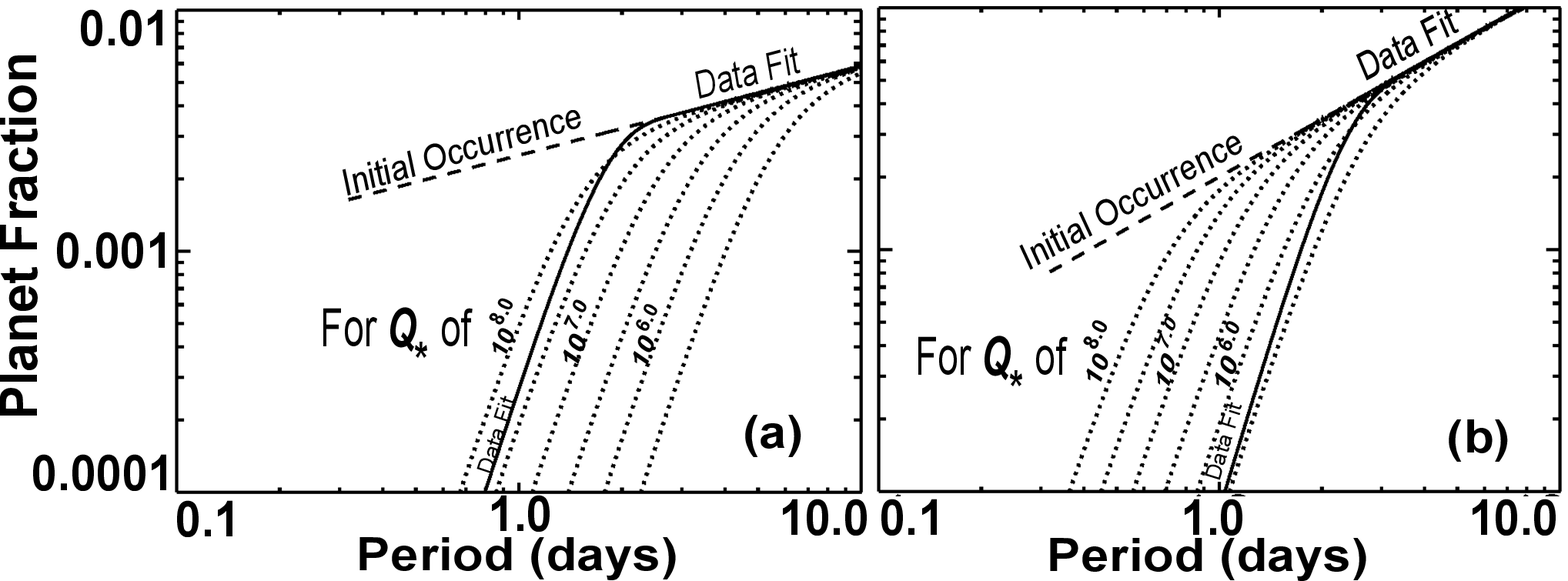}
\caption{Measured versus simulated occurrence distributions for short periods  
    for  (a) giant planets (summed for masses from $100$ to 2000 $M_\Earth$ 
    and radii from 8 to 32 $R_\Earth$) 
    and (b) medium planets (summed for masses from $10$ to 100 $M_\Earth$ and
    and radii from 4 to 8 $R_\Earth$).
    Fits to Kepler occurrence data from H12 %\citet{how12} 
    are compared to calculated distributions for 
    a range of tidal dissipation strengths from weak to strong.  %$\log(Q^{\prime}_{\ast})$  
    The left dotted-line curve is for a weak dissipation of $Q^{\prime}_{\ast}$ of $10^{8.5}$
%    ($Q^{\prime}_{\ast}$ expresses the inverse of the strength of the dissipation, so lower values
%    represent stronger dissipation), 
    with each next line representing a factor of $10^{0.5}$ stronger dissipation, 
    up to a dissipation strength of $Q^{\prime}_{\ast}$ of $10^{6.5}$.
    The simulated distributions were calculated starting from an 
    initial occurrence distribution with no inner drop off, shown as a dashed line, and
     were summed for a representative range of system ages. %cmts in FeEccFlowApJ_Text.tex
}
\label{fig:distLMo-c}       % Distribution Large Med Obs vs Calc
\end{figure}

\begin{figure}
\centering
        \begin{subfigure}[ht]{0.32\textwidth}
                \centering
                \includegraphics[width=\textwidth]{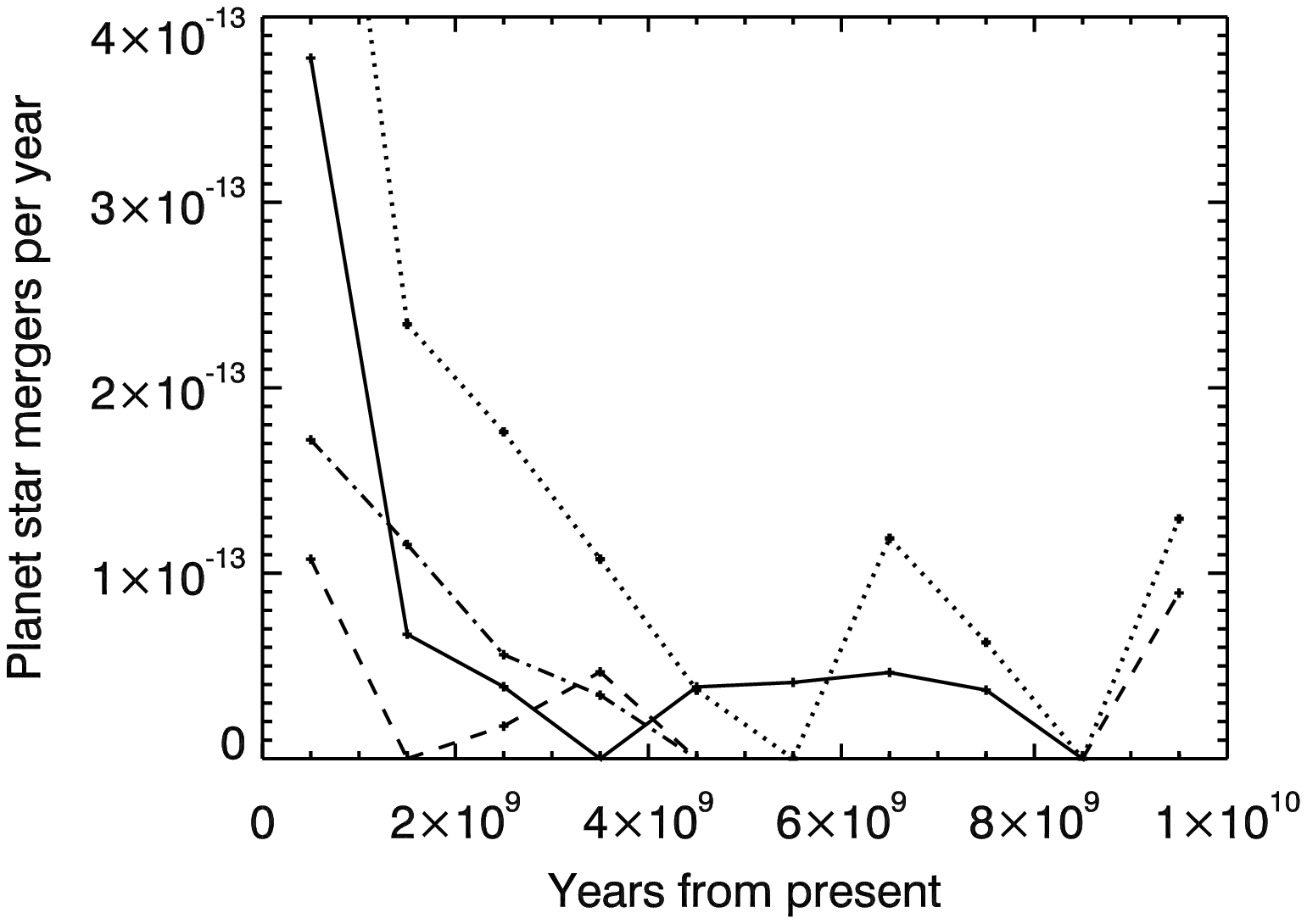}
                \caption{Infall rate of 8-16 $R_\Earth$ planets.}
                \label{fig:ratefordata_aJup}
        \end{subfigure}
        \begin{subfigure}[ht]{0.32\textwidth}
                \centering
                \includegraphics[width=\textwidth]{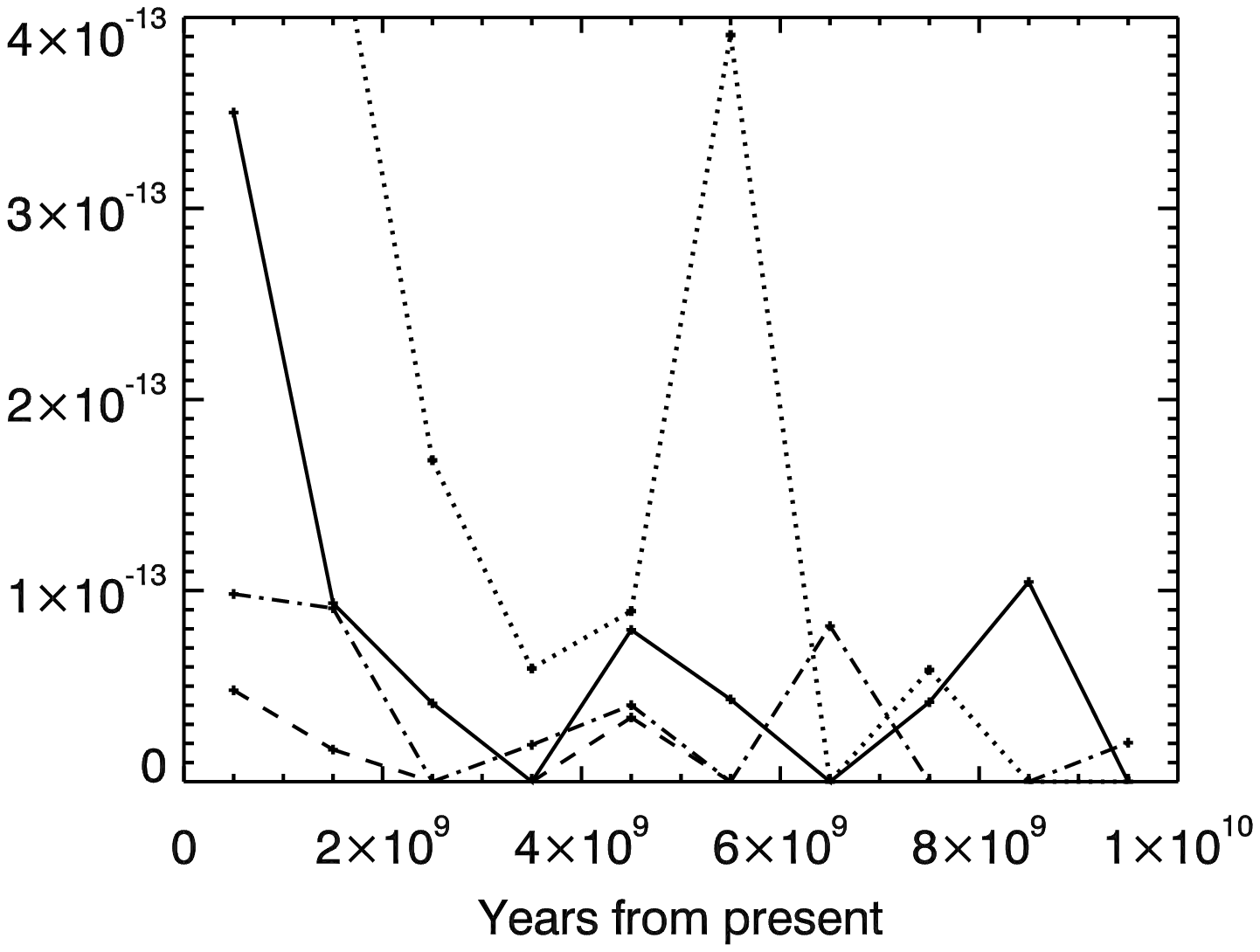}
                \caption{Infall rate of 4-8 $R_\Earth$ planets.}
                \label{fig:ratefordata_bNep}
        \end{subfigure}
        \begin{subfigure}[ht]{0.32\textwidth}
                \centering
                \includegraphics[width=\textwidth]{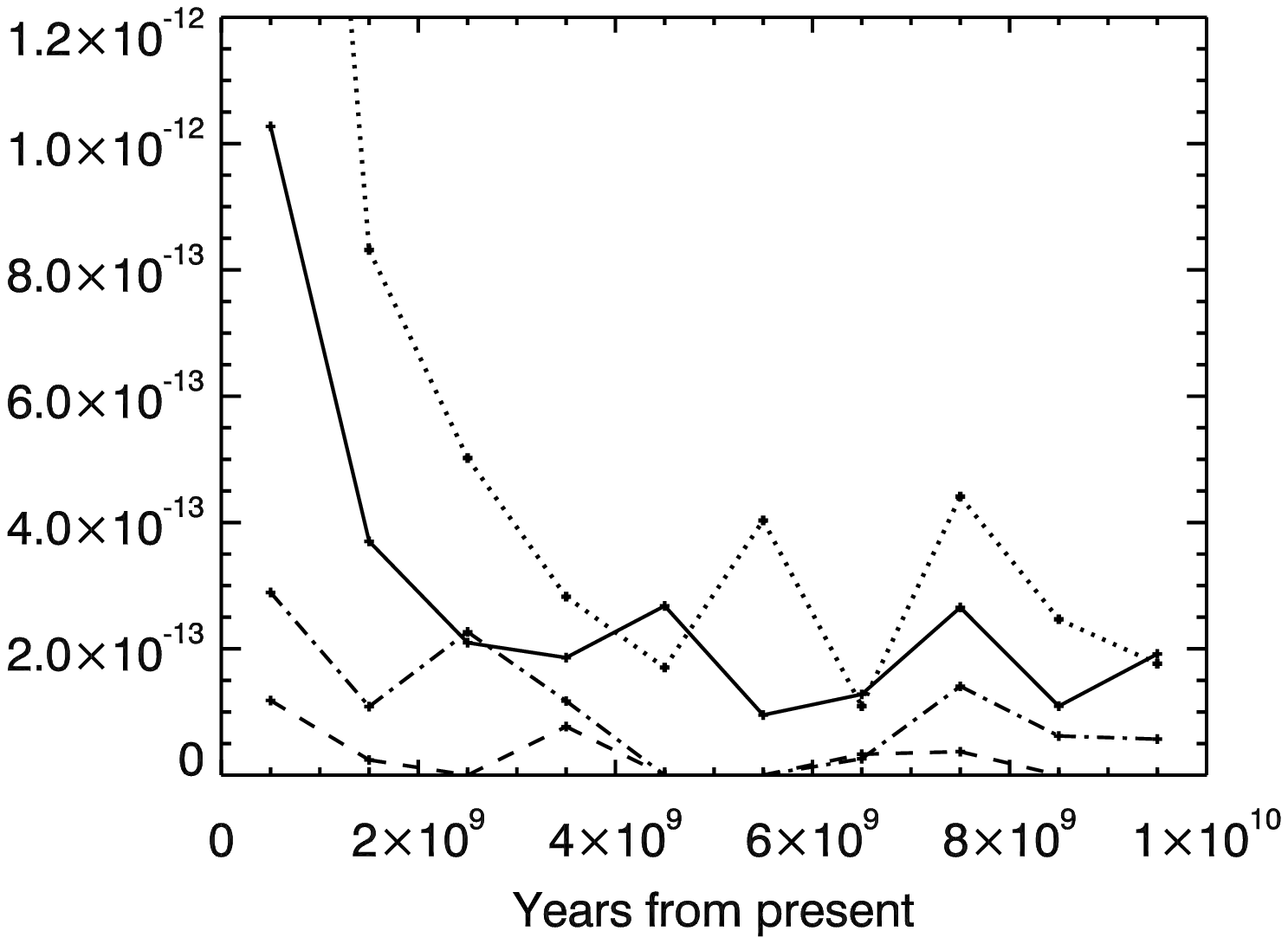}
                \caption{Infall rate of 2-4 $R_\Earth$ planets.}
                \label{fig:ratefordata_cSpe}
        \end{subfigure}
\caption{Future planet/star mergers calculated for actual Kepler candidates which are weighted 
  to give a rate that can be compared to the future merger rate presented in T13b. %\citep{tay13b}
  %Figure~\ref{fig:rateforfit}.
  %These are calculated for the same range of tidal dissipation strengths used Figure~\ref{fig:rateforfit}.
  We show connected points for dissipation values of $Q^{\prime}_{\ast}$  
  values of $10^{6.5}$ (dotted, top), with each lower line representing a tidal dissipation 
  weaker by a factor of $10^{0.5}$, down to a strength of $Q^{\prime}_{\ast}$ of $10^{8.5}$. %, for 10 Gyr.
  Though the use of individual data points makes these results more noisy, the results are 
  consistent with those from using the fit. 
  %The appearance of peaks at  5 to 8 gigayears shows 
  that the tidal dissipation is unlikely to be as strong as $Q^{\prime}_{\ast}$ of $10^{6.5}$
  as this would result in an unphysical increase in merger rate when planets beyond the fall-off 
  migrate into the star.
 }
\label{fig:ratefordata} %Using var of ~/Dropbox/idl_db/migreal/runmigkep.pro, or seriesofplots_migkep.pro, or
\end{figure}       % perhaps runmigkep_Q6789scale6.pro, all based on migkep.pro, from 2012 Aug

% stellar [Fe/H] %(hereafter [Fe/H]$_{\ast}$).

\section{More giant than medium planets at shortest periods: Weak tides or migration?}
\label{sec:MoreGiants}
%\section{Evidence of migration}  %\label{sec:MigrationEvidence}
Kepler results as presented in Figure~\ref{fig:odistrib4rangnormed} 
show more giant planets than medium planets at the shortest periods,
even though there are more medium planets further out.
Tidal migration is faster for more massive planets, so T12b and T13b %\citet{tay12b,tay13b} 
present this as evidence that giant planets regularly migrate into the star. %\citet{tay13a,tay13b} 
finds that for a tidal dissipation constant of $Q^{\prime}_{\ast} \sim 10^6$ 
that roughly than $10^{-12}$
planets per star per year would be needed to maintain the observed distribution of planets
migrating into the star; if  $Q^{\prime}_{\ast} \sim 10^7$, only a few times $10^{-13}$ planets would be required.

\citet{ham09,hel09} present how finding a planet of as short a period as that of WASP-18 is
unlikely given how soon it will merge with the star, unless it will not tidally migrate into the
star as fast as currently thought.
\citet{pen12} interpret the occurrence distribution to give a value of $Q^{\prime}_{\ast}$ 
much higher than $10^7$, but this distribution could also be interpreted as an excess due to
lower $Q^{\prime}_{\ast}$ with the short period planets continually repopulated by a flow.

Inward tidal migration operating on a  planet occurrence distribution that is initially
as a power law will produce a fall off with a power index of 13/3 (T12a,T12b). % \citep{tay12a,tay12b}.
This is close to the power index of the fall off found by H12.
%\citet[hereafter H12]{how12}. %, who fit a more uniform subset of Kepler planet candidates 
%to a ``Cutoff Power Law Model'' function.
However, the limited number of planets causes large uncertainties in the H12 fits, 
leaving it uncertain whether the power index is close to 13/3 by chance.
%More data would help resolve this uncertainty.

%\subsection{Repopulation and flow rate}
%\label{subsec:flowrate}    
An inward flow of highly eccentric planets was postulated by 
\citet{soc12}, but paucity of high eccentricity planets was seen by \citet{daw12}.
However, the  expected numbers of planets expected by \citet{soc12} is only a
few in the single Kepler field. Statistics from many fields could resolve this discrepancy. 

%\subsection{Two populations with dramatically different occurrence distributions}
%\label{subsec:twopopulatios}    
Several reports demonstrate that planets of iron-rich and iron-poor stars form two populations, 
with evidence of more crowded formation and more scattering in the iron-rich systems.

\citet[hereafter DM13]{daw13} show correlation and that the three-day pileup is a feature of the metal rich population,
but is still low in the Kepler field.
Having fields with different populations by observing stars  in different parts of the galaxy 
would allow checking whether the pileup is dependent
on location in the galaxy. This could also allow checking 
on whether the pileup is affected not only by metallicity but whether 
as H12 suggested it could be dependent on the age profile which changes as a function of distance from
the galactic plane.

Several authors find evidence of two populations of planets, with a more crowded population 
likely characterized by more giant planet migration. DM13 and T13b %\citet{daw13} %\cite{tay13b} 
find that the distribution of iron-rich and iron-poor systems have dramatically different
patterns that are likely associated with different amounts of scattering in the two populations.
These observations were 
prompted by the discovery that planet orbit eccentricity is correlated with stellar [Fe/H] 
(T12b and DM13). %\citep{tay12b,daw13}. 
\citet{xie13} find two populations of planet systems: closely versus sparsely packed, 
with the closely packed systems characterized by more transit timing variations
than sparsely packed systems.

Finding more planets is the best way to explore these abundance-dependent 
patterns. %The radial velocity method is now finding planets at longer periods.
With all the patterns that have been found at longer periods providing evidence 
for inward migration playing a major role in planetary systems evolution,
it is important to obtain better detail in the shortest period regions.

This sets up a debate with the previously preferred explanation for the discovery by 
ground surveys of planets at periods so short that finding them this close to the star should be very unlikely
if the expected strength of tidal dissipation in stars is to be believed. \citet{ham09} and others explain that 
it should be unlikely to have found number planets that appear
to be in the last small fraction of their existence before being made to tidally migrate into the star.
%He explains that the few explanations for their presence are hard to believe.
%%%\citep{pen12} use the distribution of giant planets from ground-based searches
%%%to determine that tidal dissipation in the star is most likely weaker than $Q^{\prime}_{\ast} \sim 10^7$,
%%%a value weaker than found from studies of binary stars \citep{mei05}.

However, the fit by H12 %\citet{how12} 
to the distribution of medium planets found by Kepler when compared to the 
distribution due to tidal dissipation 
as shown by Figure~\ref{fig:distLMo-c} from T12b and T13b % \citet{tay12b,tay13b} 
shows that $Q^{\prime}_{\ast}$ need not be any weaker than $10^7$.
We see in Figure~\ref{fig:distLMo-c} that the power index found by H12 %\citet{how12} 
corresponds well to the power index produced by tidal infall of the planet.
%in H12 have a power index consistent with tidal infall. 
The tidal migration into the star is fast enough in the falloff region even under weak tidal dissipation 
that the position of the falloff is not much dependent on the initial distribution
(T12b,T13b). %\citep{tay12b,tay13b}.
The biggest problem with interpreting the location of the falloff and the power index
is the uncertainty of the values from the fit of H12 %\citet{how12} 
due to the low numbers of planets available.

We show the occurrence distributions of giant, medium, super-earth radii range planets
in Figures~\ref{fig:odistrib4rangnormed}, which is normalized for transit probability and the number
of stars surveyed. To show that many of the important features in these distributions rely on counts of small
numbers, we show in Figure~\ref{fig:odistrib4rangraw} what the actual unnormalized 
counts for Figures~\ref{fig:odistrib4rangnormed} are. The falloff due to tidal dissipation 
is still poorly defined -- the fit of %\citet{how12} 
has a fall-off power index close to the value expected from tidal migration of 13/3 (T12b,T13b) %\citep{tay12b,tay13b} 
but has high uncertainty. The differences between the medium and giant planets,
and the location of the giant planet fall-off at shorter periods than the fall-off of medium planets,
rely on only a few counts. %We show these features to be worth studying with larger planet counts.

We use Kepler planet data to compare what the future infall rate of 
giant versus medium planets would be as a function of tidal dissipation strength. 
We find the rate of planets required to maintain an infalling population of shortest-period planets 
is  less than $10^{-12}$ planets per star per year (T13a,T13b), %\citet{tay13a,tay13b},  %still
a rate that would not unreasonably depopulate the reservoir of more distant planets.
This picture is still rough, however, due to not enough statistics.  
In T13b, we compare using a smooth fit to using the data directly. 
The result from T13b using the H12 fit
shows the kind of better results that could be obtained with more statistics.
The current result is uncertain due to the large uncertainty in the values of the fit found by H12.
When using the data directly, as in shown in T13b and reproduced in Figure~\ref{fig:ratefordata}, %\citep{tay13b},
%Figure~\ref{fig:rateforfit}, 
it can be clearly seen how uncertain this result is.
The presently available data for the occurrence distribution 
still allows T13b to make a case that weak tidal migration cannot maintain
a continuous infall of giant planets into the star without being
supplemented by a flow of tidally migrating giant planets.  % noisy result of using the actual Kepler 

%\citet{tay12b,tay13a,tay13b} find the explanation that these planets have undergone an event sending them migrating
%towards the star requires 
%These planets have been made
%to migrate inwards towards the star.
%Though authors had favored the explanation of weaker stellar tides, 
%\citet{tay12b,tay13a,tay13b} show 
%Planets in the shortest-period orbits
%have such a short amount of time ...
%For there to be these few planets at the closest periods 

%\subsection{Patterns in iron abundance-dependent occurrence distributions}
%\label{subsec:patterns}

%We note that that 15-day period planets could also be found using two 15 day 
%observations separated by a period of time selected to find longer period planets. 

\section{Conclusions}
\label{sec:conclusions}

More clearly measuring the shortest period distribution of giant and medium planets will show if there is
significant migration of giant planets into the star, or if the stellar tidal dissipation is weaker for
planet mass than for stellar mass companions. Such a measurement will also allow establishing
the relationship between the stellar tidal dissipation constant $Q^{\prime}_{\ast}$
and the rate of planet migration into the star. Finding more of the shortest period planets will 
prepare baseline measurements of transit times that will in not too many years enable either measuring or placing
a strict upper limit on the change of transit times, which will allow either determination of $Q^{\prime}_{\ast}$
or a strict upper limit on its maximum strength.

The HBP survey makes the best use of what Kepler is optimized for: finding planets in short periods.
%The importance of finding planets in periods corresponding to the habitable zone has been removed.
The transit method is best suited to finding the shortest period planets. Planets found by Kepler
do not suffer the effects of observing only at night, so the resulting counts are more robuts than 
counts of planets found by ground-based surveys.
Kepler was designed to find planets, and it still will do well finding planets, 
even if it does not find not as small of planets as it could before.

Studying the innermost occurrence distribution will not only allow studying tidal migration, it will
allow study of why are many of the short period giants planets inflated. It would find out
if planets sometimes become smaller due to atmospheric blowoff.
Separating out these three effects requires better statistics than those from one Kepler field.
More planets can be found by moving to new fields.
Short period planets are found much more rapidly than longer period planets. Only a dedicated
space-based survey can provide planet occurrence statistics that do not have 
the hard-to-quantify biases of ground-based surveys.
Many other proposed uses of Kepler will also propose surveying multiple fields, so the HBP mission
is compatible with other projects that require periods on the order of one month.
Surveying many fields for as many planets as possible will provide an important baseline for TESS and other studies.

This white paper gives reasons why statistics of planets in the range of one to several days has special
value. These statistics will enable further understanding the end result of tidal migration of planets.
Learning the rate at which planets migrate into the star is an important parameter for 
determining the rate of planet scattering.

%\section{Acknowledgments}
%\label{sec:acknowledgments}

%.tex}
 %\setcounter{page}{1}
 % ~/Dropbox/talks_db/2013planetloss/2013KeplerWhitePaper/HotBigPlanetsKepler_Refs.tex 
%%  Put into reference orderfrom alphabetic version which was saved as alpha file first.
% Also see ~/Dropbox/talks_db/2012planetDistributionPapers/referencesCompiled.doc 
%%  ~/Dropbox/talks_db/2013planetloss/2013feEccCorrPaper/FeEccFlowApJ_Refs.tex 
%%%%%%% REFERENCES -- no limit

% this should include only items referenced in the project description
% it is not a bibliography of related reading.

% Each reference must include the names of all authors (in the same
% sequence in which they appear in the publication), the article and 
% journal title, book title, volume number, page numbers, and year of 
% publication. If the document is available electronically, the website 
% address also should be identified

%..tex}
 % ~/Dropbox/talks_db/2013planetloss/2013KeplerWhitePaper/HotBigPlanetsKepler_Appdx.tex 
%%%%%%% REFERENCES -- no limit

% this should include only items referenced in the project description
% it is not a bibliography of related reading.

% Each reference must include the names of all authors (in the same
% sequence in which they appear in the publication), the article and 
% journal title, book title, volume number, page numbers, and year of 
% publication. If the document is available electronically, the website 
% address also should be identified

\begin{appendix}{} 

\section{APPENDIX A: Tidal migration equations}
\label{sec:appendixTidMigEqns}        

Tidal migration due to tides on the star is proportional to the planets mass.
Once planet orbits are circularized, tidal migration faster for more massive planets. 

%Struggling with getting eqn from \citet{jac09}, from fallingPlanets2010.tex 
 %~/LocalBup/talks/00fallingPaper/fallingPlanets2010.tex 

Tidal migration equations of Jackson:

\begin{align}
\label{eqn:dadtjac09}
\frac{1}{a} \frac{da}{dt}= 
- \left[ 
 \frac{63}{2} (G M^3_{\ast})^{1/2} \frac{R^5_{p}}{Q^{\prime}_{p} M_{p}}e^2
+ \frac{9}{2} (G/M_{\ast})^{1/2} \frac{R^5_{\ast} M_p}{Q^{\prime}_{\ast}}
 \left( 1+\frac{57}{4} e^2 \right) \right] a^{-13/2} 
\end{align}

For the eccentricity evolution, we use the equation of Jackson et al. (2009),

\begin{align}\label{dedtjac09}
\frac{1}{e} \frac{de}{dt}=                  % \nonumber and &  for two rows (see 2010ppr)
- \left[ 
  \frac{63}{4} (G M^3_{\ast})^{1/2} \frac{R^5_{p}}{Q^{\prime}_{p} M_{p}}
%\\ %&        													         %& for two rows (see 2010ppr)
 + \frac{225}{16} (G/M_{\ast})^{1/2} \frac{R^5_{\ast} M_p}{Q^{\prime}_{\ast}}
  \right] a^{-13/2} 
\end{align}

\end{appendix}

\end{document}